\documentclass{acmsiggraph}

\usepackage{booktabs}

\usepackage{amssymb}
\usepackage{amsmath}

\usepackage{xspace}

\usepackage{textcomp}
\usepackage{xcolor}
\usepackage{float}
\usepackage{multirow}

\usepackage{algorithm}
\usepackage{algpseudocode}

\renewcommand{\comment}[1]{}
\input{Author/Commands.tex}

\usepackage{xspace}

\usepackage{textcomp}
\usepackage{color}
\usepackage{float}
\usepackage{wrapfig}
\usepackage{multirow}

\TOGonlineid{25}

\CopyrightYear{2019}
\setcopyright{acmcopyright}
\conferenceinfo{CONFERENCE PROGRAM NAME}{MONTH, DAY, and YEAR}
\isbn{THIS-IS-A-SAMPLE-ISBN}\acmPrice{\$15.00}
\doi{http://doi.acm.org/THIS/IS/A/SAMPLE/DOI}

\title{Animating an Autonomous 3D Talking Avatar}

\comment{
\author{Dominik Borer}
\affiliation{
  \institution{ETH Zurich \& Disney Research}
}
\author{Dominic Lutz}
\affiliation{
  \institution{Disney Research}
}

\author{Robert W. Sumner}
\affiliation{
  \institution{ETH Zurich \& Disney Research}
}
\author{Martin Guay}
\affiliation{
  \institution{Disney Research}
}}

\author{Dominik Borer, Dominic Lutz, Martin Guay}

\pdfauthor{Dominik Borer}

\begin{document}

\teaser{
\includegraphics[width=0.7\linewidth]{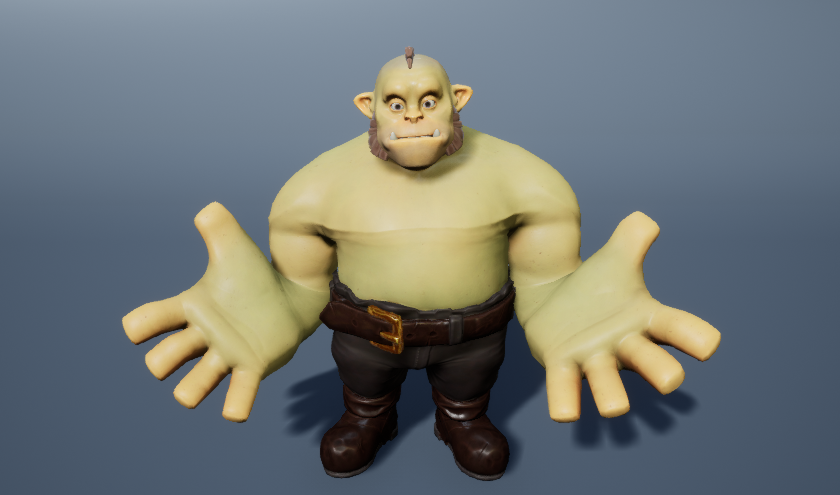}
\caption*{}
\label{Teaser}
}

\maketitle

\begin{abstract}  

One of the main challenges with embodying a conversational agent is annotating how and when motions can be played and composed together in real-time, without any visual artifact. The inherent problem is to do so---for a large amount of motions---without introducing mistakes in the annotation. To our knowledge, there is no automatic method that can process animations and automatically label actions and compatibility between them. In practice, a state machine, where clips are the actions, is created manually by setting connections between the states with the timing parameters for these connections. Authoring this state machine for a large amount of motions leads to a visual overflow, and increases the amount of possible mistakes. In consequence, conversational agent embodiments are left with little variations and quickly become repetitive. In this paper, we address this problem with a compact taxonomy of chit chat behaviors, that we can utilize to simplify and partially automate the graph authoring process. We measured the time required to label actions of an embodiment using our simple interface, compared to the standard state machine interface in \textit{Unreal Engine}, and found that our approach is 7 times faster. We believe that our labeling approach could be a path to automated labeling: once a sub-set of motions are labeled (using our interface), we could learn a prediction that could attribute a label to new clips---allowing to really scale up virtual agent embodiments.



\end{abstract}

%
%
\comment{
}


%
%


\keywordlist

\conceptlist

%
%

\newcommand{\figStateMachineUnrolled}{	
	\begin{figure}[!htb]
		\centering
		\includegraphics[width=1.0\linewidth]{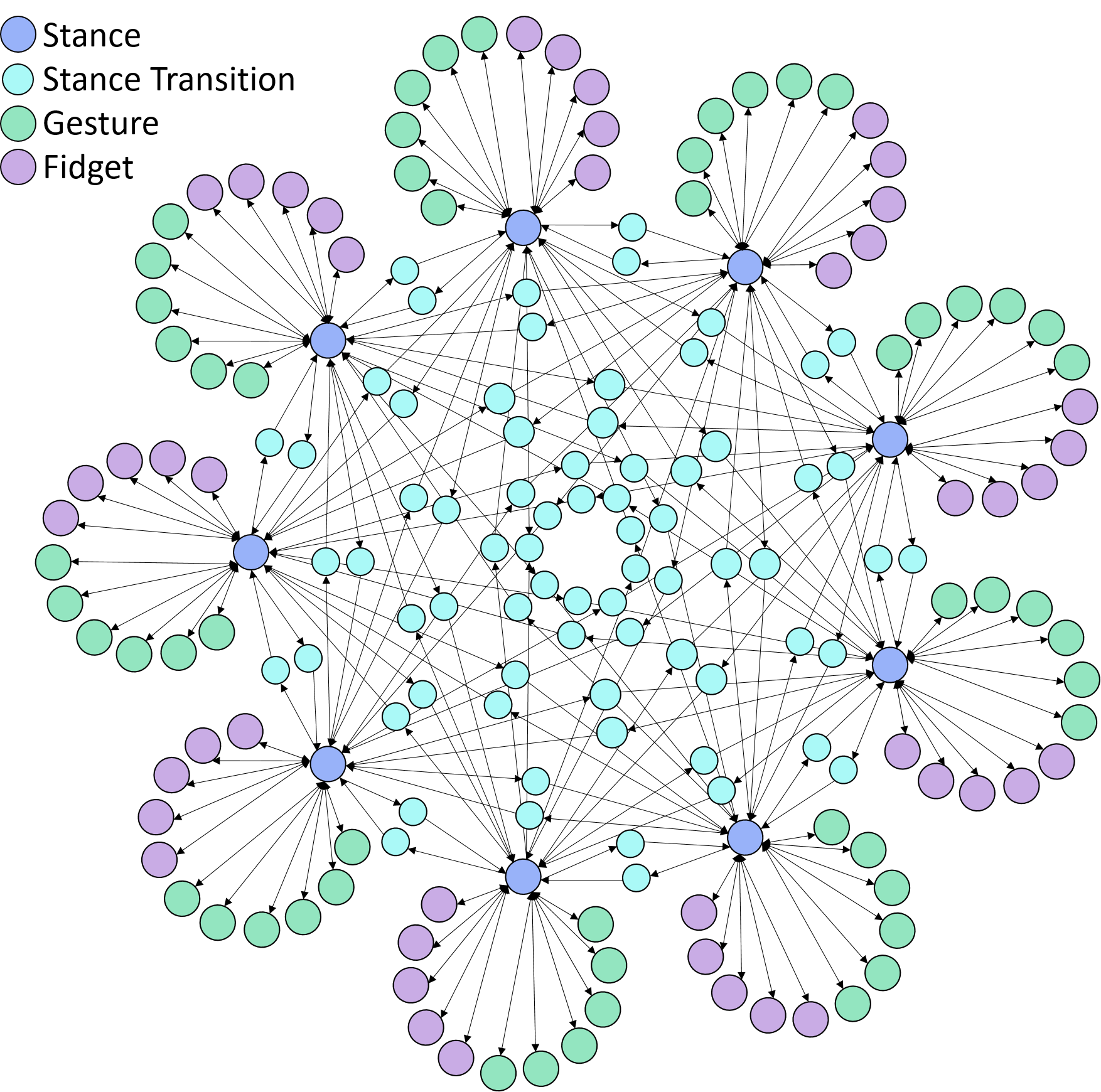}
		\caption[State Machine Unrolled]{Explicitly creating the animation state machine with all connectivity information quickly becomes complex and prone to errors. Since the logic for all transitions is the same (up to some motion-specific parameters), there is a lot of redundant information, which we can reduce to a simple interface as shown in \figref{fig:MetaNodes}.}
		\label{fig:StateMachineUnrolled}
	\end{figure}
}

\newcommand{\figMetaNodes}{	
	\begin{figure}[!htb]
		\centering
		\includegraphics[width=1.0\linewidth]{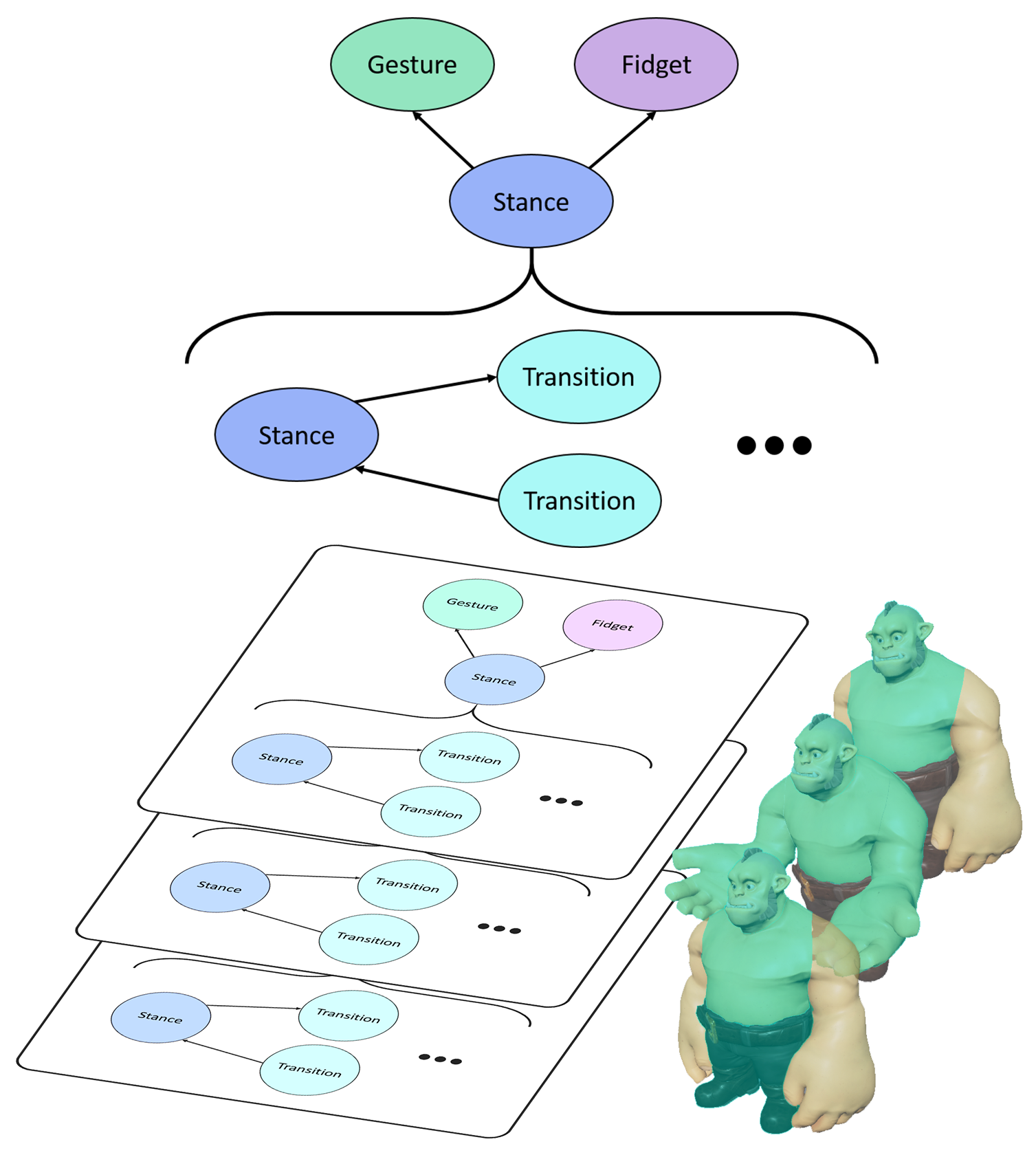}	
		\caption[Drag and Drop Meta Nodes]{We can observe the self-similarities in the full graph (\figref{fig:StateMachineUnrolled}), and generalize the connections between meta nodes coresponding to our taxonomy (stance, gestures, figets and stance transitions). This abstract graph, we call the meta-graph, naturally leads to a simplified interface for users to simply drag-and-drop animation clips over the nodes, as described in \secref{sec:Interface} and shown in \figref{fig:Interface}. }
		\label{fig:MetaNodes}
	\end{figure}
}

\newcommand{\figIdleActionStateMachine}{	
	\begin{figure}[!htb]
		\centering
		\includegraphics[width=0.5\linewidth]{Author/Images/GenericStateMachine.png}
		\caption[Idle Action State Machine]{All possible motion sequences can be abstracted to idle-action-idle-action sequences, which can be reproduced using this basice idle-action state machine. Two action states are required to avoid artificial gaps and visual artifacts for subsequent actions.}
		\label{fig:IdleActionStateMachine}
	\end{figure}
}

\newcommand{\figLayers}{	
	\begin{figure}[!htb]
		\centering
		\includegraphics[width=0.45\linewidth]{Author/Images/BorkLayers.png}
		\caption[Layers]{This figure is dangerously confusing: at a superficial level it looks like an overide type of masking... To combine multiple motions we split the character up into five layers--head, spine, legs and left and right arm.}
		\label{fig:Layers}
	\end{figure}
}

\newcommand{\figAdditiveBasePose}{	
	\begin{figure}[!htb]
		\centering
		\includegraphics[width=0.5\linewidth]{Author/Images/BasePose.png}
		\caption[Additive Base Pose]{The base pose $\basePose$ used to compute and composite the additive offsets $\additivePose{}$.}
		\label{fig:AdditiveBasePose}
	\end{figure}
}

\newcommand{\figAllLayers}{	
	\begin{figure}[!htb]
		\centering
		\includegraphics[width=0.9\linewidth]{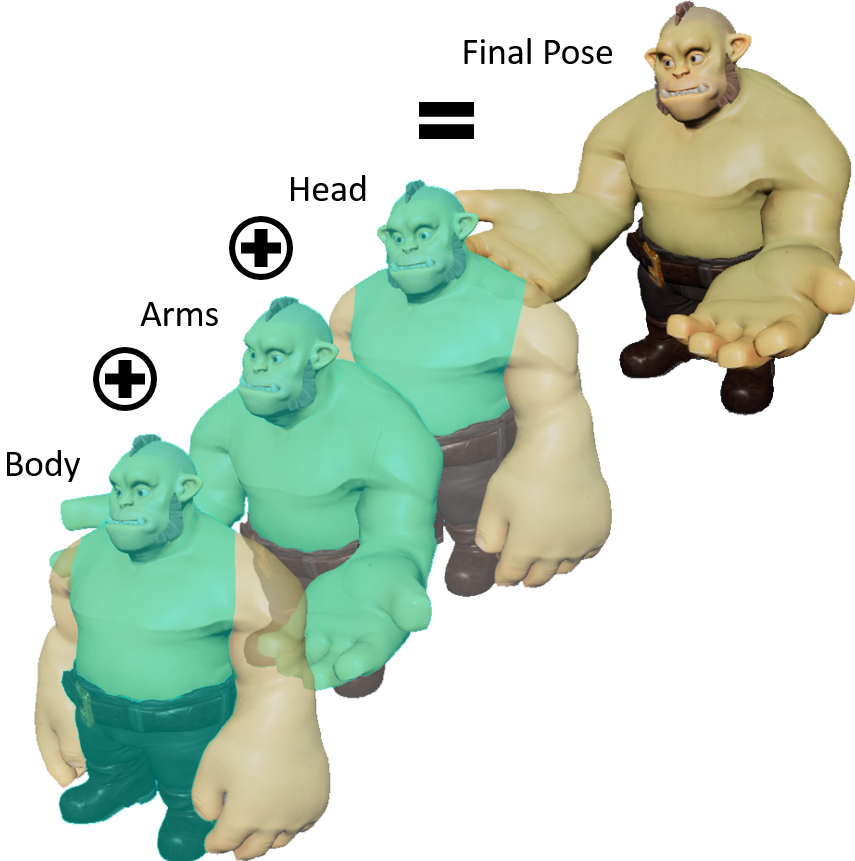}
		\caption[All Layers]{The animations triggered on the different layers \textit{head}, \textit{arms}, and \textit{body}, are composed additively to build the final pose. The influences of the layers on the different body parts are highlighted in green.}
		\label{fig:AllLayers}
	\end{figure}
}

\newcommand{\figInterface}{	
	\begin{figure}[!htb]
		\centering
		\includegraphics[width=1.0\linewidth]{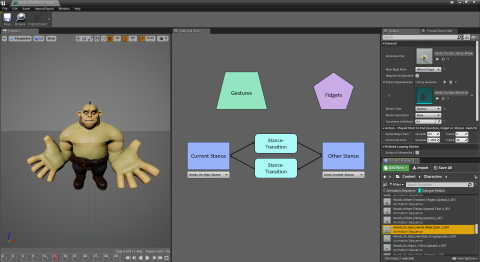}
		\caption[Interface]{To add new motions, users drag-and-drop animation clips onto the corresponding meta nodes. Additional properties such as layer masks and timings can be specified in the properties panel.}
		\label{fig:Interface}
	\end{figure}
}

\newcommand{\figActionRequestsToMetaAction}{	
	\begin{figure}[!htb]
		\centering
		\includegraphics[width=1.0\linewidth]{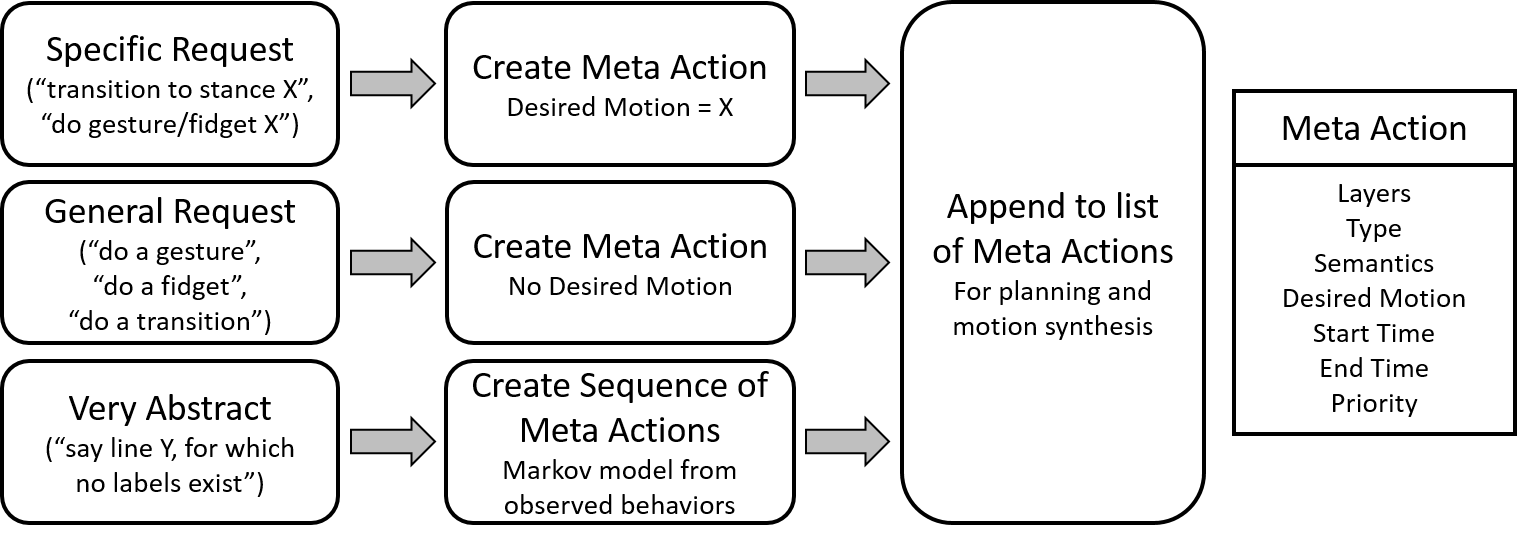}
		\caption[Action Requests To Meta Action]{Action requests at various levels of abstraction are converted into \textit{meta actions}, a unified representation containing all required data for further planning (\secref{sec:TemporalReplanning}) and motion synthesis (\secref{sec:SamplingSPecificActions}).}
		\label{fig:ActionRequestsToMetaAction}
	\end{figure}
}

\newcommand{\figMetaActionAndPlan}{	
	\begin{figure}[!htb]
		\centering
		\subfigure {\label{fig:MetaAction}\includegraphics[width=0.2\linewidth]{Author/Images/MetaAction.png}}
		\subfigure {\label{fig:MetaPlan}\includegraphics[width=0.2\linewidth]{Author/Images/MetaPlan.png}}		
		\caption[Meta Action and Meta Plan]{The properties of a \textit{meta action} and the structure of the \textit{meta plan}.}
		\label{fig:MetaActionAndPlan}
	\end{figure}
}

\newcommand{\figTemporalReplanning}{	
	\begin{figure}[!htb]
		\centering
		\includegraphics[width=1.0\linewidth]{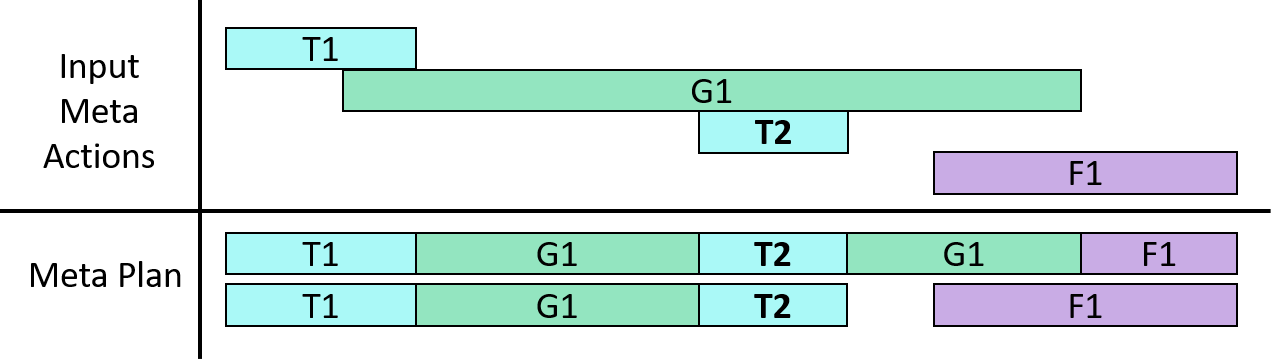}
		\caption[Temporal Replanning]{Through temporal replanning the meta actions are converted into a \textit{meta plan}, a conflict-free sequence of meta actions, by adjusting the timings greedily according to starttime and priority. For long actions interrupted by shorter actions of higher priority we either continue with the remaining part after the cut or discard it.}
		\label{fig:TemporalReplanning}
	\end{figure}
}

\newcommand{\figSamplingSpecificActions}{	
	\begin{figure}[!htb]
		\centering
		\includegraphics[width=0.9\linewidth]{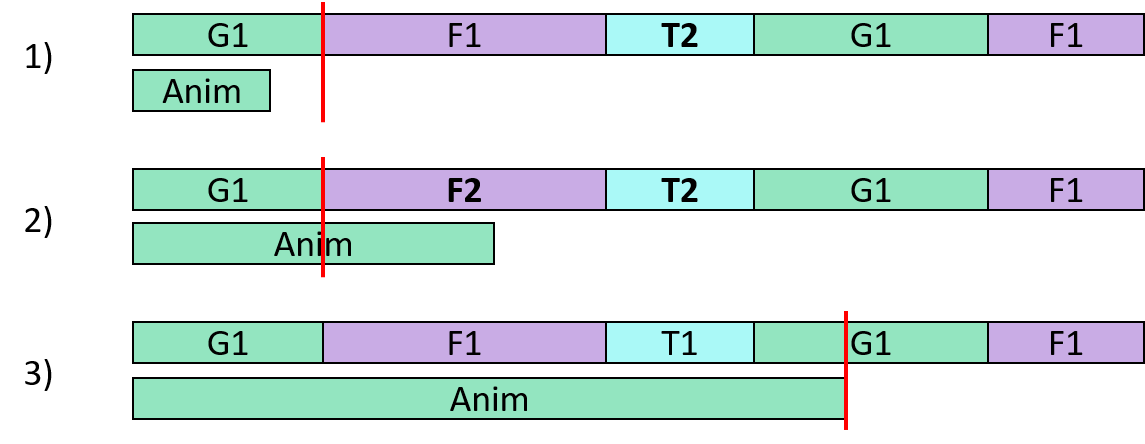}
		\caption[Sample Specific Actions]{To synthesize a motion, animations are sampled according to the meta action properties. We distinguish three cases during this process: 1) If the animation ends before the meta action, we continue at the begin of the next meta action. 2) The animation has to end before a meta action with higher priority. If none is compatible, the meta action is discarded. 3) Meta actions completely overlapped by the aniamtion are discarded and we continue at the end of the animation.}
		\label{fig:SamplingSpecificActions}
	\end{figure}
}

\newcommand{\figConversationBorkSteven}{	
\begin{figure}[!htb]
	\centering
	\includegraphics[width=1.0\linewidth]{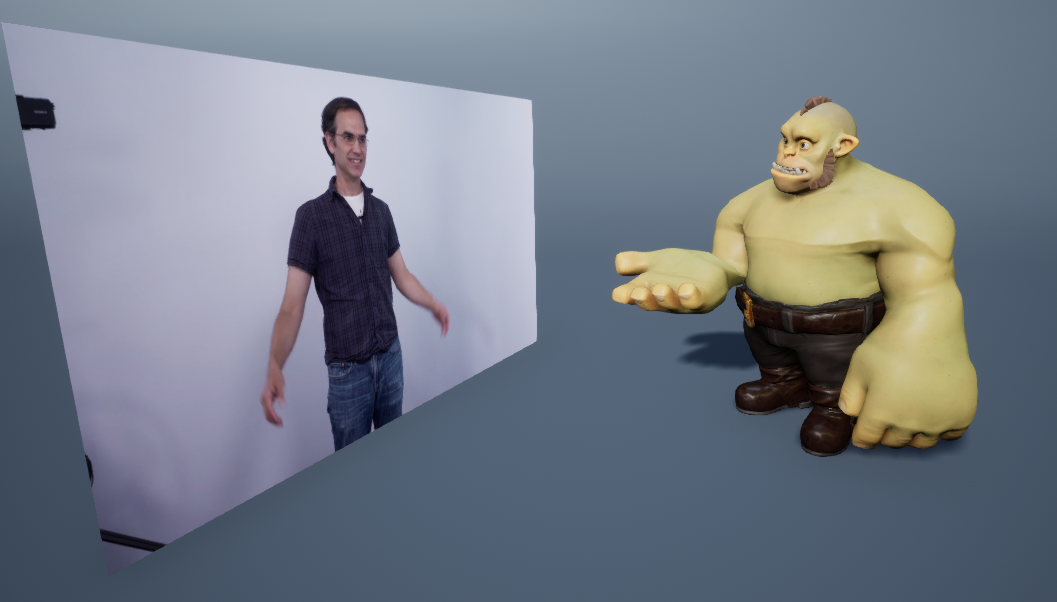}
	\caption[Conversation Bork Steven]{Face-to-face conversation between a user and the virtual character. To generate the action sequences, we annotated audio as shown in \figref{fig:ElanLabelling}.}
	\label{fig:ConversationBorkSteven}
\end{figure}
}

\newcommand{\figMotionVariety}{	
\begin{figure}[!htb]
	\centering
	\includegraphics[width=1.0\linewidth]{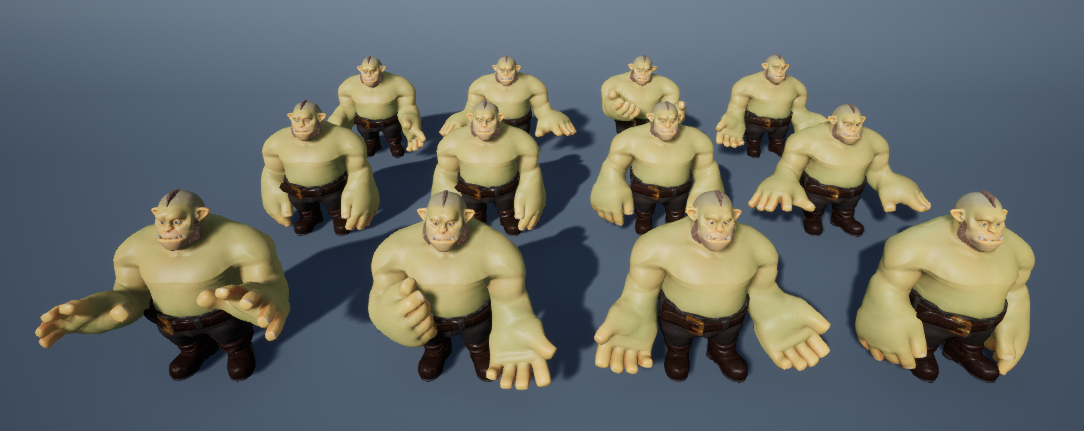}
	\caption[Variety of Motions]{Due to the probabilistic nature of the motion synthesis, for the same abstract action sequence different performances emerge. This includes using different animations as well as different timing.}
	\label{fig:MotionVariety}
\end{figure}
}

\newcommand{\figElanLabelling}{	
	\begin{figure}[!htb]
		\centering
		\includegraphics[width=1.0\linewidth]{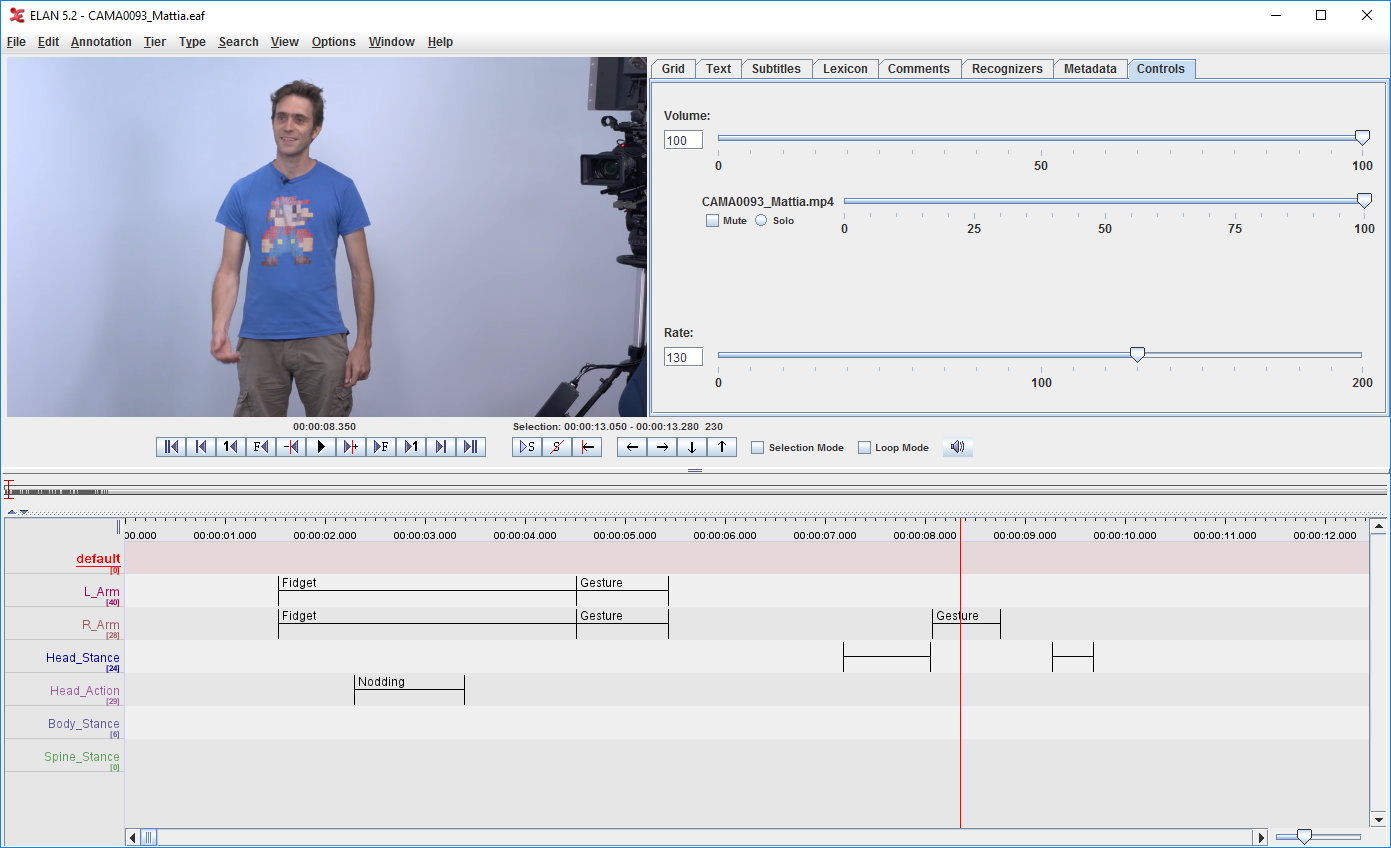}
		\caption[Elan Labelling]{To create action sequences for casual conversations, we annotated the recorded dyadic conversations using the ELAN software \cite{Elan}. These annotations include abstract actions for different body part layers, which include \text{head stance}, \textit{head action}, \textit{left arm}, \textit{right arm} and \text{legs}. These annotations can then be used either directly as actions (\figref{fig:ConversationBorkSteven}), or to learn a Markov model (\secref{sec:MarkovModel}).}
		\label{fig:ElanLabelling}
	\end{figure}
}

\newcommand{\figTransitionMatrix}{	
	\begin{figure}[!htb]
		\centering
		\includegraphics[width=1.0\linewidth]{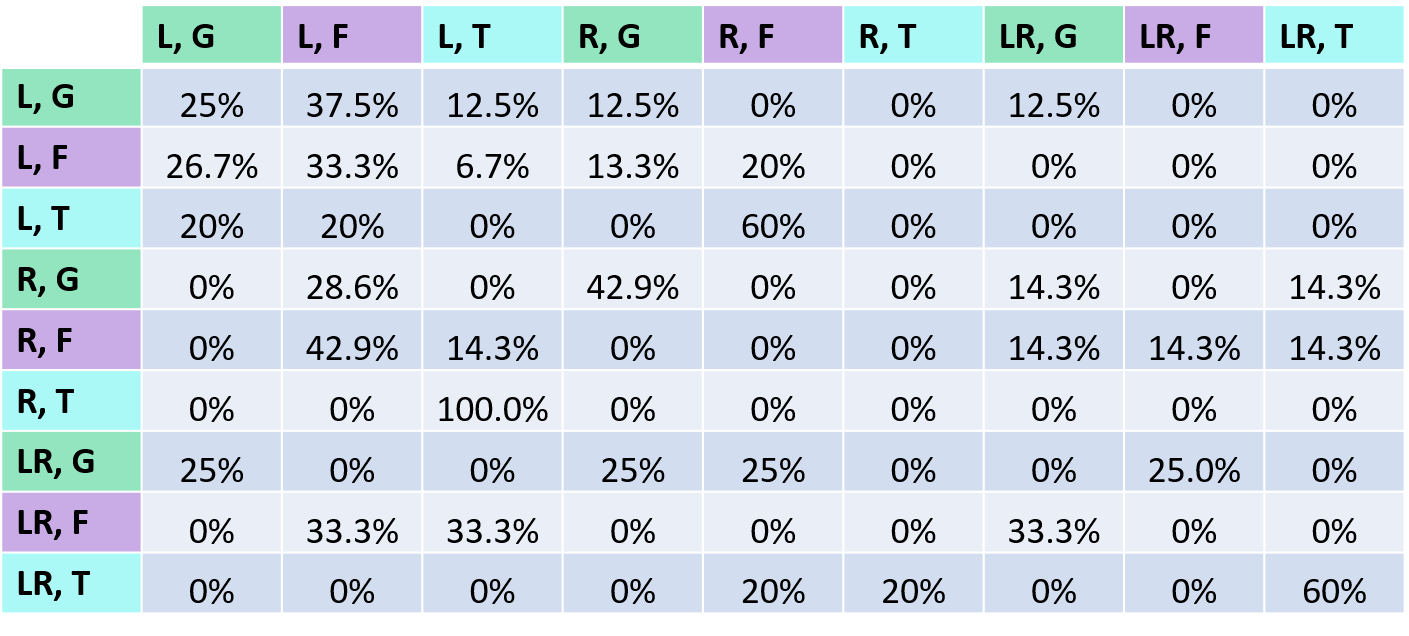}
		\caption[Transition Matrix]{Using the labeled audio, we can build a first-order Markov-model. The transition matrix, in this example for the arms, shows the probability between the different states of the model, which consist of the left~hand~(L), right~hand~(R) and both~hands~(LR) together with the actions gesture~(G), fidget~(F) and transition~(T). }
		\label{fig:TransitionMatrix}
	\end{figure}
}

\newcommand{\figGeneratedSequence}{	
	\begin{figure}[!htb]
		\centering
		\includegraphics[width=1.0\linewidth]{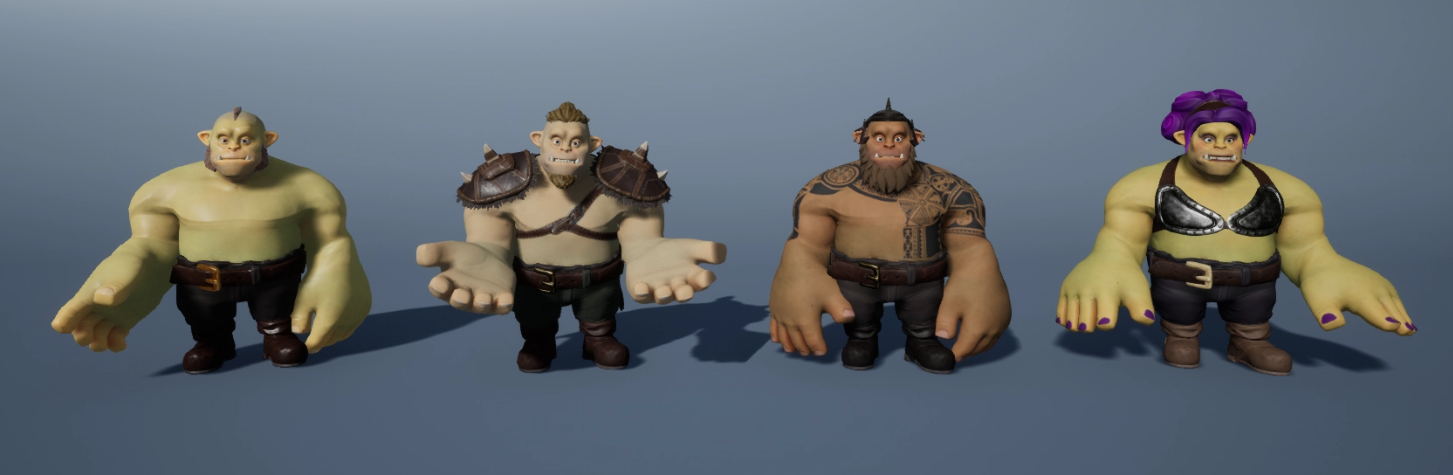}
		\caption[Generated Sequences]{Using the learned Markov model, we can generate new action sequences that look natural. Since the generated actions are still abstract, the resulting performances will vary.}
		\label{fig:GeneratedSequences}
	\end{figure}
}

\section{Introduction}
\label{sec:Introduction}

Intelligent avatars that can talk and interact with people offer a natural and friendly way to interface with autonomous systems---may they be cars, televisions, food dispensers, phones, maps, and so on. While this vision is not new, machine perception and natural language processing has evolved rapidly in recent years, causing a re-visitation of this vision by various companies, each releasing voice-based assistants such as \textit{Alexa}, \textit{Siri}, or \textit{AliGenie} and some embodied systems with digital displays such as \textit{Gatebox}, \textit{Jibo} and \textit{Baidu}.

One of the main challenges with embodying an agent is annotating how and when motions can be played and composed together in real-time, without any visual artifact. The inherent problem is to do so---for a large amount of motions---without introducing mistakes in the annotation. To our knowledge, there is no automatic method that can process animations and automatically label actions and compatibility between them. In practice, a state machine, where clips are the actions, is created manually by setting connections between the states with the timing parameters for these connections, together with other state parameters such as body part mask, clip start and end time, as well as whether the motion is looping or not. For example, consider an action that has the arm waving in the air, while a new action requests a beat gesture that starts with the arm pointing on the floor. The sudden jump to the other motion will cause a visual artifact. Hence in this case the state machine would have a transition between both gestures, and the timing would be set to the end of the first clip, forcing the embodiment to finish the first clip before playing the next (assuming they have a compatible pose at their respective extremities).

Authoring this state machine for a large amount of clips requires repetitive labour for many actions, and can grow exponentially for certain types of behaviors that require transitions between them, as shown in \figref{fig:StateMachineUnrolled}. Hence the possibilities of introducing mistakes is very high. In consequence, conversational agent embodiments are often left with little variations, and quickly become repetitive and predictable. In this paper, we address this problem with a compact taxonomy of chit chat behaviors, that we can utilize to simplify and partially automate the graph authoring process. We observed from casual dyadic conversations that people are mainly in \textit{stances}, \textit{fidgets}, \textit{gestures} and \textit{transitions (between stances)}. Stances are synonymous to idle (e.g. arm on waist, or body weight on one side), fidgets are ticks and small subtle gestures, while \textit{gestures} are more functional. The final element are the transitions between the stances such as changing the body weight to another side, or having a hand going from the waist to a shoulder.

With our taxonomy, we propose a more abstract and partially automated interface for the state machine authoring, where animation clips are dragged and dropped onto one of the action types (gesture, fidget, stance, transition), as shown in \figref{fig:MetaNodes}. We asked an animator to create animations guided by our taxonomy, and to utilize our interface to populate the graph. In total, 152 animations were created, and labeling all the motions (or authoring the graph) was 7 times faster than using the traditional interface in Unreal Engine (see our comparison discussion for more details \secref{sec:smauthoring}). We believe that our labeling approach could be a way to automated labeling via a bootstrapping phase: once a sub-set of motions are labeled (using our interface), we could learn a prediction that could attribute a label to new clips---allowing to really scale up virtual agent embodiments.

We additionally report on our best practices for embodiment decision making (\secref{sec:PlanningAndSynthesis}, which we learned during a collaborative project for embodying an agent. We found that in practice agents operate at different levels of abstraction. For example, one agent might output high-level actions such as ``say line 279'', or lower level commands such as ``nod here''. To cope with different levels of abstraction, we designed a motion planner that can filter, as well as synthesize actions, if needed. Our planner takes sequences of \textit{abstract} actions and outputs a sequence of \textit{specific} actions. In the case of too few actions, the planner can generate natural sequences using probability distributions conditioned on the previous action, which we compute from video recorded human performances.

\comment{
By observing casual conversations, we found that the overall motion behavior can be broken down into \textit{stances}, \textit{transitions between stances} and actions from the stances such as \textit{gestures} and \textit{fidgets}. From an animator's perspective this gives a clear set of motions to create and from the state machine editing perspective we can predefine connections or graph primitives that can be appended to generate a dense graph without manual editing. The center piece is the stance, hence the animator selects a stance and can drag-and-drop animation clips on to the appropriate slots for the graph connectivity to be updated.
}


\figStateMachineUnrolled

\comment{
We assume in this work an embodiment that is stylized and not human-like. On the one hand, this may help avoid the uncanny valley, and on the other, it may be a design restriction where say a popular cartoon character is to serve as embodiment. The consequence is the lack of large and dense datasets for the character, prohibiting the use of a machine learning type of approach.

The second aspect to consider is the level of abstraction of the agent. A typical agent takes various sensory input such as video and audio, and seeks to output a set of action sequences for the embodiment to perform. A practical problem arises which is: at which level of abstraction should the actions be specified? Should it and can it plan at the low level of joint angles? Hence, it is useful in practice to support different levels of abstraction: to offer the highest levels of abstraction such as ``say line of text X'', as well support ``perform a \textit{beat} gesture here'', or ``head nod here'', as well as specific actions (``perform beat gesture \#13 here''), as well as controlling directly joint angles.

To support such parameterizations of the action space of the character or robot, designers typically use a \textit{state machine}, in which different states are motion clips, and the logic between which clips can be played after which is specified by connections and transition rules between the states. These are heavily used by video games to build interactive parametric character controllers.

While state machines allow any type of function approximation, in practice they quickly become complex to manipulate as the number of states (actions, animations) grows. The reason is that we can only play animations that are compatible with one another, since blending from one to the other will otherwise cause visual artifacts if only joint space interpolation is used. For example, an arm motion starting from the hip cannot be triggered when the arm is currently on the shoulder. This compatibility information is encoded into a state machine via connections that are manually specified by the designers and engineers, which often results in bugs and mistakes as the number of actions grows, as can be seen in \figref{fig:StateMachineUnrolled}.

In this paper, we solve this scalability problem with an abstraction of the casual conversation behaviors, allowing to automate a large portion of the state machine construction, and support a more compact representation for the user. By observing casual conversations, we found that the motion space can be broken down into \textit{stances}, which are close to static poses, \textit{transitions between stances}, and then actions from the stances such as \textit{gestures} (e.g. a beat hand gesture, or a head nod) and \textit{fidgets}, which are subtle motions such as ticks. This abstraction allows for a more compact and robust template interface, where animators simply drag-and-drop animation clips into one of the slots, and the graph's connectivity is automatically updated.

To cope with different levels of abstraction, we designed a motion planner that can filter, as well as synthesize actions, if needed. Our planner takes sequences of abstract actions and outputs a sequence of specific actions. In the case of too few actions, the planner can generate natural sequences using probability distributions conditioned on the previous action, which are obtained from observed behaviors from real human performances. Finally, we demonstrate the capabilities of our casual embodiment of abstract action labels, and demonstrate also the combination of \textit{abstract} together with \textit{specific} actions. To this end, we video-recorded dyadic conversations, which we hand-labeled with abstract postural and gestural actions. At runtime, our planner generates a performance of \textit{specific} actions. The probabilistic or generative nature of our planner, allows re-generating different (but similar) performances for the same abstract actions. This can also be combined with \textit{specific} actions specified by other means. We included a neural ``mimicry'' model for the body, which infers the specific \textit{stance} to take, based on a video feed from the person talking to the character.

We conducted an evaluation to measure the time saved when using our drag-and-drop semi-automatic interface for designing the embodiment graph, versus the time required to specify it manually with the standard state machine interface in \textit{Unreal Engine}. The standard path took 9 days to finalize the dense state machine due to the increasing amount of connections to author and maintain, while our approach required only 1.5 days.
}

\comment{
Contributions:
\begin{itemize}
\item An abstraction of casual conversation allowing a automatic editing of the connectivity. (Ontology: the stance, transition between stances, gestures and fidgets. (Assumes independence between layers.)
  \item A drag-and-drop interface automating graph connectivity (effectively abstracting away the details of the graph for the user). User only needs to drag-and-drop clips, and specify the \textit{begin} and \textit{end} timings.
  \item This template interface utilizes a graph composition framework which appends a basic building block to encompass the larger graph topology.
\end{itemize}
}

\comment{
Casual conversation actions \textit{ontology}:
\begin{itemize}
  \item \textit{A stance} is a permanent postural state (the body stance left or right, or the head tilt left, or the arm on the hip).
  \item A fidget or micro-gestural action, is a small action such as tick such as scratching, or a rotating the body slightly while standing, which people often do naturally.
  \item \textit{Gestures} are actions such as an arm gesture to communicate time or space structure, or a head gesture to communicate acknowledgment.
\end{itemize}

The state machine requires connections between states, or animation clips, as well as the timings in which transitions can occur, together the action type they correspond to (stance, fidget or gesture, or transition). The typical interface in a game engine is a graph in which a user must establish connections manually. In our case, the number of motions would quickly create large star shaped graphs.

We have a dense graph with transitions between all stances
a transition is a stance-switch state (animation clip) from one stance to another.
We have a dense fully connected graph (conceptually).
}

\section{Related Work}
\label{RelatedWork}

\textbf{Agency, Planning and Architecture related to Embodiment. }
Over the past years there have been many works on turning text and other modalities into actions.
The seminal work by Casell et al. on text to motion \cite{Cassell1998, Cassell1999a, Cassell1999b}, culminating in \cite{Cassell2001}, use natural language processing together with manually designed rules and heuristics to create a sequence of nonverbal behaviors from text. Similarly rule based approaches have been used to create talking head animations from chat text in online games \cite{Vilhjalmsson2004}.
The realization of the motions is often performed for the whole body, without advanced additive composition \cite{Thiebaux2008} and the interface for the action specification is often the Behavior Markup Language BML \cite{BML2006}.
Similar recent work use besides text also speech (audio) features \cite{Marsella2013}, and have been applied to a virtual therapist conducting a depression screening therapy \cite{DeVault2014}.
Other works on text to motion, build a probabilistic model of gestures for a specific speaker from annotated video \cite{Neff2008a}.

To play animation clips, often state machines are used, which can conceptually be viewed as graphs. The works \cite{Arikan2002, Kovar2002} focuses on automatically building such a graph based on similarity between poses in large data-sets. To match animations to action specifications more accurately in time, others have investigated optimal matching with dynamic programming \cite{Stone2004, Bozkurt2016}, which requires allowing re-timing of the clips via warping. In our experiments, we fonud that the warpings cause the resulting motion to look unnatural.


\bigskip\noindent\textbf{Engineering Feature-based Maps. } When the motion can be generated from a set of features, or parameters known to the agent, such as the audio (the speech), or high-level parameters such as the gaze direction, or even a part of the motion specified otherwise, such as the head orientation for gaze, then a large amount of variations can be synthesized automatically, especially when the parameter and the mapping are continuous.

Lip sync is one case where audio is analyzed and corresponding lip shape parameters are computed over time \cite{LipSyncPro, FaceFX}. In the case of the body, recent work has explored to map audio features to the trunk motion of the character---mapping the volume of the audio to the trunk and head arc \cite{Ishiguro2016}. Adding arm gestures in an override fashion from the shoulder downwards, removing essentially all spine motion from the arm's action, results in a motion that looks uncanny. Hence some map the arm motion back to the spine motion to bring back some of the missing dynamics, as described in chapter 9 of \cite{Tanenbaum2018}.


\bigskip\noindent\textbf{Machine Learning. } Mapping audio or speech to lip motion is a relatively well defined problem as strong correlations exist between the speech and mouth shape used to produce the sounds, resulting in many methods and papers on the topic, summarized in \cite{Mattheyses2015}. Researchers have pushed the envelope with speech-driven eyebrow motion \cite{Ding2013}, and more recently with speech-driven facial animation, which can produce motions in a given expression \cite{Karras2017, Taylor2017, Sadoughi2017}. In case of the body, experiments have been conducted to learn a mapping from speech to body motion, by first capturing the motion of an actor while talking \cite{Levine2009, Levine2010}. Similarly an interesting recent work seeks to map music audio to body motion performed while playing the instrument to produce the sounds \cite{Shlizerman2017}.

But generally the results are poor, perhaps due to the articulated nature of the human body, but most likely due to the lack of powerful latent structures that can model the natural gestures that accompany speech content. Recent works experimented with the idea that additional modality, such as the face of the person talking to the avatar, together with audio and transcripts, could improve the results in a deep learning setting \cite{Chu2018}.

\section{Overview}

It is not possible to simply play any motion at any point in time when using the standard joint space interpolation provided in game engines such as \textit{Unity} or \textit{Unreal Engine}. For example, consider a gesture that moves the arm up, and then a sudden request for an action that starts with the arm pointing downward. Interpolating between both motions will cause a visual jump.

To avoid these artifacts, engineers and designers encode which motion can be played when or after which other motion as connections in a state machine, where the states are the motion clips. For example, consider arm motions and following our taxonomy guideline: stances for the arms such as hands on the hips, transitions between different stances, and all the gestures and fidgets that can be played from those stances. Imagine we have 9 stances, and about 5 gestures and fidgets on each stance. We would end up with a graph that looks like \figref{fig:StateMachineUnrolled}. Authoring and maintaining these connections, requires exponential effort as the number of animations grows, and the probability of making mistakes gets inreasingly high. In consequence very few embodiments scale up to rich and highly varied conversation behaviors.
\figMetaNodes
When looking at this large graph, we can observe the self-similarities between the sub-structures of the graph. We can generalized these sub-structures into \textit{meta nodes}, as shown in \figref{fig:MetaNodes}. This generalized graph is a perfect fit for a simplified drag-and-drop interface that automates the connectivity logic, while animators focus on the motions (\secref{sec:Interface}).

To this point, we discussed arm motion, but when people are talking, they perform head motions such as nods, while doing different hand gestures and occasionally changing body postures such as shifting the weight to one side of the body. Creating all the combinations of head motions with gestures, fidgets and weight shifts leads to exponential content authoring. To cope with this complexity, we break down the motion space into body part layers (we call \textit{body}, \textit{arms}, \textit{head}), and compose them in an additive fashion (\secref{sec:AnimationComposition}). Note that our taxonomy still holds for the different body parts. Hence, the graph on each layer is the same, and we explain in \secref{sec:Interface} how the interface for specifying the motions on each layer works.

Finally, we tackle the problem that different agents can operate at different levels of abstraction, which makes it difficult for the embodiment to be used in practice. An agent might specify action types such as ``beat'' gesture at a given time, or have specific gestures it wants to play. Another agent might want to simply send dialogue lines without any actions. By modeling the probability distributions of the gestures, we designed a planner that can translate action sequences at different levels of abstraction into specific action sequences without conflicts (\secref{sec:PlanningAndSynthesis}). In other words, our planner can generate plausible action sequences when none provided, and can avoid repetitive gestures when more abstract actions are specified. That said, we begin by describing the user interface for specifying the graph.

\section{User Interface}
\label{sec:Interface}
Based on the generalized meta graph shown in \figref{fig:MetaNodes} we devise a simple drag-and-drop interface to automate the connectivity logic between the animations as illustrated in \figref{fig:Interface}.

First the current stance, which is the central piece, is specified either by selecting an existing one or by drag-and-dropping a new one. Gestures and fidgets are then simply added by drag-and-dropping them onto the propper nodes. Similarly transitions are added, but first the other stance has to be selected. Any additional properties such as layer masks and timing together with optional details such as semantic information or base likelihood are specified in the properties panel on the right.
\figInterface
By drag-and-dropping animations we fill the graph data structure, which is a multi-index map, where for each stance and meta-node type (taxonomy element type), we have a list of compatible motions. Additionally for each transition motion we also store in which stance it ends. Next we detail the inner workings of the state machine playback, together with the additive layer composition.

\section{Animation Composition}
\label{sec:AnimationComposition}

Creating all combinations of head motions with gestures, fidgets and weight shifts is not feasible and we therefore break the motion space down into body layers. Specifically we decompose the motions into three layers: \textit{body}, \textit{arms}, and \textit{head}. Simply composing the layers by masking, results in robotic and uncanny motions, because the dynamics for other body parts is lost. Instead, by composing the motions for the different layers additively, we can maintain this dynamics. From analyzing the motions we observed the following influences between the three layers (body, arms, head) and the four body parts (legs, spine, arms and head):
\begin{alignat*}{5}
	& body & \ \rightarrow & \ head, \ && spine, \ && legs \\
	& arms & \ \rightarrow & \ head, \ && spine, \ && arms \\
	& head & \ \rightarrow & \ head, \ && spine && 	
\end{alignat*}
The final pose for each body part is thus composed as the additive combination of these three layers as visualized in \figref{fig:AllLayers}. The base pose used for this is a neutral pose, where the character has the hands at his side.
\figAllLayers
To create an additive animation, all joint transforms $Q_{anim}$ are converted into offsets $\additivePose{}$ relative to those of the base pose $\basePose$ such that $Q_{anim} = \basePose \addOp \additivePose{}$, where $\addOp$ is the additive operator (multiplication for orientations and addition for translations). The final poses for all layers are then composed as:
\begin{align*}
	& \finalHeadPose = \basePose \addOp \additiveWeightBody \additiveBody \addOp \additiveWeightArms \additiveArms \addOp \additiveWeightHead \additiveHead \,, \\	
	& \finalSpinePose = \basePose \addOp \additiveWeightBody \additiveBody \addOp \additiveWeightArms \additiveArms \addOp \additiveWeightHead \additiveHead  \,, \\	
	& \finalLegsPose = \basePose \addOp \additiveWeightBody \additiveBody \,, \\
	& \finalArmsPose = \basePose \addOp \additiveWeightArms \additiveArms \,,
\end{align*}
where $\additiveWeight{}$ are influence weights, $\additivePose{}$ additive offsets, $\pose$ final poses and $\{\headIndex \,, \armsIndex \,, \bodyIndex \}$ refer to the \textit{head}, \textit{arms} and \textit{body} layers. For the influence weights we used $\additiveWeight{} = 1.0$ for all layers. Despite not normalizing the weights, we did not observe any visual artifacts. We assume that with more extreme motions, it might be necessary to perform some normalization, possibly through optimization.


A final note: to make the character more alive we apply a breathing motion that affects spine, head and arms (slight shoulder motion), lip-sync for the speech and life-like eye motion, further detailed in \apxref{sec:EyeMotion}. These are applied in the same additive fashion as the other motions.

Now that we have a lively character, we look at how to combine it with the agents actions.

\comment{
\subsection{Avoiding an Explicit Graph}

While the meta nodes shown in \figref{fig:MetaNodes} allow creating a large graph from a simple interface, we observed that it was possible to completely avoid explicitly representing the graph. When the character is performing a stance, it is a subtle stance animation that loops, while in \textit{transitions}, \textit{gestures} and \textit{fidgets} states, the animation plays from start to end. Hence we can abstract these to a binary state: either \textit{idle} or \textit{action}. In consequence, any action sequence performed can be broken down into a sequence of idle-action-idle-action states. Instead of explicitly encoding all of the connections between states, we can define a basic idle-action state machine, shown in \figref{fig:IdleActionStateMachine}, and use this building block at run-time.


\figIdleActionStateMachine

During a state transition the animations are cross-faded over some time window. During that time we cannot swap out animations, but always having to wait for the blend to finish may create unnecessary time in idle between subsequent actions. Having two action sates allows to immediately transition to the next action. We also observed that stance transitions are usually long enough such that this problem does not occur and thus one idle state is enough.
}

\section{Planning and Motion Synthesis}
\label{sec:PlanningAndSynthesis}

In practice it may be hard to predict at which level of abstraction an agent is going to operate: will it be at the level of \textit{"say line X"}, or will it be at the level of \textit{"lift right hand index finger by 10 degrees"}? To support the various levels of abstraction, we designed a \textit{planner} that first converts streams of actions from the agent, into \textit{meta actions}, which are a unified representation containing all required meta data. 

The meta actions (shown in \figref{fig:ActionRequestsToMetaAction}) contain an abstract action field with values that match our taxonomy: "gesture, fidget, stance transition", as well as a specific action field, together with an additional property field for dimensions such as postivie - negative, and finally a timing field. This tupple of abstract and specific allows to accept both, more specific actions, as well as more abstract actions.

\figActionRequestsToMetaAction

The next ingredient to our planner are probilistic models of the gestures and actions. In the event of unusually absent actions, or when an agent acts only at a very high level such as "utter line X", the planner generates sequences of \textit{meta actions} using a sequential probability distribution. We compute a Markov model from \textit{labeled} real world casual conversations, which we detail further in \secref{sec:MarkovModel}.

Once we have the sequences of \textit{meta actions}, our planner turns them into spefic actions: first by resolving temporal confilcts, then by making sure gestures don't repeat using a probability distribution of the actions conditioned on past actions.

\subsection{Temporal Replanning}
\label{sec:TemporalReplanning}

Besides potential conflicts among the input meta actions, there may be conflicts with an already existing \textit{meta plan}, which is a conflict-free sequence of meta actions as shown in \figref{fig:TemporalReplanning}, created at an earlier timestep. From the existing plan we first recover all meta actions that start after the currently active action and merge them with the inputs. Through temporal replanning we then create a meta plan by adjusting the timings of the meta actions in a greedy way according to start time and action priority as illustrated in \figref{fig:TemporalReplanning}. 

\figTemporalReplanning

\subsection{Sampling Specific Actions} 
\label{sec:SamplingSPecificActions}

In the last stage, the meta plan is translated into a sequence of specific actions (animation clips). For a given meta action, we first retrieve all potential candidates by matching meta action properties against the specific motion data set. The possible actions may introduce conflicts temporally, as the three cases shown in \figref{fig:SamplingSpecificActions}. We remove these from the possible actions.

From the final set of possible actions, we sample according to a distribution that is conditioned on past actions. We assume here that all actions are uniformly distributed, unless specified differently through the interface, but modulate the distribution to reduce the probability of having recurrent motions. We also consider the three cases illustrated in \figref{fig:SamplingSpecificActions} in our sample selection.

\figSamplingSpecificActions

Given a count of past actions $\{ c_i \}$ for the $n$ candidates, the probability of picking candidate $k$ is:
\[ p_k = \frac{\frac{1}{\max(c_k, 1)}}{\sum_{i = 1}^{n} \frac{1}{\max(c_i, 1)}} \]
When an animation has been used, its counter is updated as:
\[ c_k = c_k + \alpha \,, \]
where $\alpha$ is a weight determining how quickly the probability decreases the more often a motion is used. We used $\alpha = 4$.

Now that we have the agent connected to the animation system, we continue with experiments and results.

\section{Experiments and Results}
\label{sec:Results}

\comment{
\review{A center-piece of the paper is the comparison between using our interface, and using unreal directly. We need to describe this comparison. }
}

One of the benefits of our taxonomy is to effectively guide animators and artists on which motions to create: stances, stance transitions, gestures and fidgets. However, it is quite challenging to evaluate these benefits, as creating the animations for an embodiment each time represents months of labour. 

That said, we could leverage our taxonomy into automation, and the time required to populate the sate machine with actions is much lower than with the standard state machine utilities in a game engine. In practice, the state machine increases in complexity as the number of motions increases, \textit{and} requires engineering skills to find problems (see \secref{sec:smauthoring} for additional discussion on authoring state machines). To confirm this reality, we conducted an evaluation by measuring the time taken for an animator to populate the state machine using our interface, and the time taken using the standard state machine utilities in \textit{Unreal Engine}.

Our animator has three years experience with the \textit{Maya} software, and has no engineering background. We first explained the interface with an example of 2 stances, 1 gesture, 1 fidget and 2 stance transitions, each explaining the layers for the body parts and motion-specific parameters. It required a total of 8 hours spread over two days (1.5 days of work) to populate the state machine with our 152 animation clips. Using directly Unreal Engine, our animator had a 1 week tutorial on authoring state machines with frequent help required to accomplish tasks. After this 1 week tutorial, our animator started creating a state machine manually. It took 63 hours (8.5 days) of work to finalize an equivalent state machine. The resulting motion holds several jumps and artifacts and it holds many more mistakes.

We conducted several experiments to demonstrates the look and feel of our embodiment, which can be viewed in our accompanying video. However, we did additional work to accomodate agent actions at different levels of abstraction. To demonstrate this capability we conducted 3 experiments. First we hand-labeled videos of dyadic conversation with meta labels (see \secref{sec:PlanningAndSynthesis}) and show that the embodiment can generate variations of the same meta action sequences (\secref{sec:conversationlabeling}). A second experiment demonstrates the combination of two different streams of actions: a meta-actions stream together with specific actions coming from an interactive mimicry module (\secref{sec:Mimicry}). Finally, we learned probability distributions of the meta actions and show that we can generate plausible performances in the absence of actions from the agent (\secref{sec:MarkovModel}).



\figConversationBorkSteven
\figMotionVariety

\subsection{Discussion on State Machine Authoring}
\label{sec:smauthoring}

When authoring a state machine in Unity or Unreal Engine (or any modern game engine), motions are added by creating a new state. Several properties such as looping or not, start and end times, and so on, have to be specified. To use this state, all the connections to other states (clips) have to be established, which as the state machine grows, becomes more and more demanding. The growing number of connections shown on screen quickly leads to a \textit{visual overflow}, making it highly prone to errors. For example, forgetting a connection or connecting to the wrong state leads to visual artifacts in the performance. Additionally, a transition requires transition logic such as the time at which it is possible to transition, and the type of blend between clips. This process is quite redundant as actions of the same meta class, all behave the same in terms of transitions. In contrast, our interface has transition logic pre-defined for the meta action classes: stance, stance transitions and the gesture/fidget categories. 




\subsection{Annotating Dyadic Conversations (Video)}
\label{sec:conversationlabeling}
We video recorded two hours of dyadic conversation: 2 times 1 hour, each with 2 subjects. A list of casual topics was given to each subject in case conversation runs dry. We captured the upper body above the knees, as can be seen in our accompanying video and \figref{fig:ElanLabelling}.

To annotate the videos with our abstract action labels (stance, stance transition, gesture or fidget), we used the \textit{ELAN} software \cite{Elan}, shown in \figref{fig:ElanLabelling}. To capture the different combinations of actions, we created labels for multiple body layers: two layers for the head (one for \textit{stance} and one for \textit{action} such as nodding), one layer for each arm, and one layer for the body, legs and spine. For the arms, the labels included \textit{gesture}, \textit{fidget} and \textit{stance-transition}. For the head action, the labels included \textit{nodding} and \textit{shaking}. And for the head stance and body layers, the labels only included \textit{stance-transition}. 

\figElanLabelling

Given these sequences of abstract action labels, we can synthesize a performance by feeding them to our planner. Since the labels remain abstract, we can synthesize different performances for the same input sequence. \figref{fig:MotionVariety} shows an example of various hand gestures, synthesized from the same input. 

\subsection{Modeling Action Distributions with a Markov Model} 
\label{sec:MarkovModel}
In the absence of labels, the embodiment can generate plausible conversation motion by leveraging the label distributions. We build a \textit{first-order Markov-model} from the labeled video and sample the distribution to generate actions sequences. The possible states of the Markov model are the abstract actions \textit{gesture}, \textit{fidget} and \textit{stance transition}. When additional semantic information is required such \textit{positive} for nodding, we can either randomly pick a semantic label when landing in a gesture state, based on its distribution in the annotations, or we can include it into the Markov model as additional states (\textit{gesture-without-semantics}, \textit{positive-gesture}, \textit{negative-gesture} etc.). Then at generation time, we take the current Markov state and sample the next state based on the observed transition probabilities. \figref{fig:TransitionMatrix} shows a possible transition matrix for the different Markov states and \figref{fig:GeneratedSequences} shows different generated performances. The full performances can be seen in the accompanying video.

\figTransitionMatrix
\figGeneratedSequence

\subsection{Video-based Mimicry}
\label{sec:Mimicry}
To demonstrate combining two streams of actions, one abstract and one specific, we took as input a mimicry module that reads the pose of a person in a video and classifies the stance, together with the stream of actions for the arm and head gestures. Given the detected stances, we pass an action of the form ``transition to the detected stance'' to the planner. If the stance is different from the current stance, the planner then finds the corresponding transition to trigger.
Similarly the mimicry module also detectes the orientation of the spine and head and then passes an action of the form ``shift weight to the detected side'' or ``tilt head to the detected side''. Again, if the requested side for weight shift or head is different from the current, the planner finds the corresponding motion to trigger.

\section{Conclusion}
\label{sec:Conclusion}

We introduced a compact action taxonomy for chit chat embodiment, together with a fast interface for labelling motion clips \textit{at a large scale} (152 actions in our experiment).  Users require no engineering skills to connect hundreds of motions, with regard to their body part and action type (stance, gesture, fidget and stance transition): they simply drag-and-drop clips onto the corresponding sheet and adjust timings in a side window. Compared to the standard state machine authoring system, our interface avoids many human mistakes and requires significantly less time (7 times faster by our measurement). Holding a large number of motions and variations yields a richer and more natural looking performances, as can be seen in our accompanying video. Finally, our fast labeling interface could lead to additional automation: given a large set of labelled actions, we could investigate training a predictor to label new similar motions---allowing to fully automatically add motions to the agent's embodiment.



\appendix
\section{Life-like eye motion}
\label{sec:EyeMotion}

Eyes do not simply blink at regular intervals, but perform subtle horizontal and vertical \textit{eye saccade} \cite{Meur2015}. We include eye saccade using hand-crafted animations for the eye joints, which we compose additively onto the base eye motion possibly coming from other animations.

Then we felt that the eye lids look uncanny. We observed that with real humans, there is a close relationship between the position of the eyelids, and where we look at. For example, when looking down, eyelids are slightly closed. Generally we can say that the upper and lower eyelids follow the upper and lower bounds of the iris respectively. This is implemented by mapping the eye rotation around the horizontal axis to the blendshape weights for the eyelids.

During conversations we usually do not stare at the other person all the time, but rather avoid the gaze from time to time by looking at some other point, before returning back to the person. To add this to our character, we implemented a smooth look-at system to control the view direction of the character. At somewhat regular intervals, we set a random direction at an offset from the subjects' face. The character smoothly looks at the other direction for a small period of time, and then smoothly comes back to its default direction. 



\bibliographystyle{acmsiggraph}
\bibliography{Avatars}  

\end{document}